\documentclass[a4paper,11pt]{article}
\usepackage{jheppub} 

\title{De Sitter geometric inflation from dynamical singularities}
 
\author[a,b]{Adolfo Cisterna,}
\author[c,d]{Nicolás Grandi,}
\author[e]{Julio Oliva}

\affiliation[a]{Sede Esmeralda, Universidad de Tarapacá,\\ Avenida Luis Emilio Recabarren 2477, Iquique, Chile}
\affiliation[b]{Institute of Theoretical Physics, Faculty of Mathematics and Physics,
Charles University, V Hole{\v s}ovi{\v c}k{\' a}ch 2, 180 00 Praha 8, Czech Republic }
\affiliation[c]{Departamento de Física, Universidad Nacional de La Plata, \\ Casilla de Correos 67, La Plata, Argentina.}
\affiliation[d]{Instituto de Física La Plata, Consejo Nacional de Investigaciones Científicas y Técnicas, \\ Diagonal 113 esquina 63, La Plata, Argentina.}

\affiliation[e]{Departamento de Física, Universidad de Concepción, \\ Casilla, 160-C, Concepción, Chile.}

\emailAdd{adolfo.cisterna.r@mail.pucv.cl}
\emailAdd{grandi@fisica.unlp.edu.ar}
\emailAdd{juoliva@udec.cl}
\abstract{Within the framework of geometric inflation, where the Friedmann equation is modified to incorporate an infinite series of higher curvature corrections, we describe the emergence of a de Sitter inflationary phase near the poles of an arbitrary dynamical function. Our analysis is quite general and does not depend on any specific choice of cosmological dynamics.}
\begin{document}  
\maketitle
\vspace{-0.8cm}
\section{Introduction}
It was recently identified that an $\mathcal{R}^3$ correction to General Relativity (GR) provides a framework for geometric inflation \cite{Arciniega:2018fxj}. This specific combination of cubic curvature invariants is determined by the requirement that it leads to second-order field equations for a Friedmann-Lemaître-Robertson-Walker (FLRW) cosmological ansatz. Such a combination has previously been identified in the context of black holes in higher-curvature gravity theories \cite{Bueno:2016xff, Hennigar:2017ego}. Quartic and quintic invariants with similar properties in the FLRW ansatz were determined in \cite{Cisterna:2018tgx}, where it was also shown that scalar cosmological perturbations satisfy field equations of second order in time. Assuming the existence of a curvature invariant at any order, leading to second-order equations in the FLRW ansatz, Ref. \cite{Arciniega:2018tnn} explored various scenarios leading to an early inflationary expansion for the Universe, smoothly transitioning into a radiation-dominated era, and subsequently to the $\Lambda$CDM model (see also \cite{Arciniega:2019oxa, Arciniega:2020pcy} for further explorations). The existence of such models to all curvature orders was recently demonstrated by Moreno and Murcia in \cite{Moreno:2023arp}, who also proved that the equations for the scalar perturbations of the FLRW ansatz can be made second order in time.

On the other hand, in \cite{Cisterna:2018tgx} we established a connection between the aforementioned combinations of higher curvature terms and five-dimensional theories known as Quasitopological Gravities \cite{Oliva:2010eb, Myers:2010ru, Dehghani:2011vu, Cisterna:2017umf, Bueno:2019ycr}. These theories exhibit remarkably simple properties when applied to spherically symmetric backgrounds. It was recently observed in \cite{Bueno:2024dgm} that, in the context of effective field theory, any combination of polynomial curvature invariants 
can be re-written as one of the Quasitopological class after a suitable field redefinition. Then, evaluating the field equations on a spherically symmetric ansatz, the authors were able to resum the series of contributions coming from the infinite number of higher curvature corrections, leading to different models admiting black holes with a de Sitter core, namely, regular black holes purely supported by higher curvature contributions and with no matter source whatsoever.

The aforementioned works put on firm ground the exploration of four-dimensional cosmologies in sensible theories containing an arbitrary number of higher curvature corrections. This work aims to achieve this objective. Remarkably, we 
show that 
regardless of the precise values of the coefficient in front of each curvature term, 
the early evolution of the Universe is generically dictated by an inflationary de Sitter epoch, which can be smoothly connected with a radiation dominated era\footnote{We refer to \cite{HZ} and \cite{CNunez} for the existence of non-perturbative de Sitter solutions in String Theory to all orders in $\alpha'$ coming from $O(d,d)$ invariance restrictions, and \cite{Liu} for a thorough exploration of these solutions both in the Einstein and String frame.}. 

\section{Basic {geometric inflation} setup}
We describe a $3+1$ dimensional homogeneous and isotropic cosmology using the Friedman-Lemaître-Robertson-Walker (FLRW) metric: 
\begin{equation}
ds^2=-dt^2+a^2(t)\left(\frac{dr^2}{1-k\,r^2}+r^2d\Omega_2^2\right)\,.
\label{eq:metric.FRW}
\end{equation}
As usual, $a(t)$ is the scale factor and $k = 0, \pm 1$ represents the spatial curvature.

We focus on a gravitational dynamics where the equation of motion for the scale factor $a(t)$ is of second-order and autonomous. This allows us to write a first integral in the form of a Friedmann-like equation \cite{Arciniega:2019oxa, Arciniega:2018tnn, Cisterna:2018tgx}
\begin{equation}
P\left(H^2+\frac {k}{a^2}\right)=\frac\kappa3\left(\frac {\rho_r^0}{a^4}+\frac{\rho_m^0}{a^3}\right)+\frac\Lambda 3\,.
\label{eq:Friedmann}
\end{equation}
%
Here $H = \dot{a}/a$ is the Hubble parameter, and the right-hand side includes the energy densities of radiation and matter, and a cosmological constant. The function $P(x)$ encodes the specific gravitational theory, with $x = H^2 + k/a^2$ representing the spacetime curvature. 

The explicit form of $P(x)$ arises from the coefficients in the curvature expansion of the gravitational action, specific examples can be found in the literature \cite{
Arciniega:2018fxj, Cisterna:2018tgx, Arciniega:2018tnn, Arciniega:2019oxa, Arciniega:2020pcy, Edelstein:2020nhg}.
In the present note we will keep $P(x)$ as general as possible.  We can reabsorb any constant contribution to $P(x)$ in the cosmological constant $\Lambda$ in order to set $P(0)=0$. Since we want to recover Einstein gravity at small curvatures, we impose $P(x)\approx x$ for small $x$, where any proportionality constant has been reabsorbed in the definition of $\kappa$.

As the cosmology evolves following equation \eqref{eq:Friedmann}, the spacetime curvature $x$ changes. We assume that at any particular point of the evolution $x=\bar x$ the function $P(x)$ shows one of two  possible behaviours: 
\begin{enumerate}
    \item It can be regular $P(x)\sim (x-\bar x)^p$ with $p\in\mathbb{Z}_+$.
    \item It can have a pole of a given finite order $P(x)\sim (x-\bar x)^{-p}$ with $p\in\mathbb{Z}_+$.
\end{enumerate}
Depending on which of the two possibilities is realized, the cosmology around that particular point shows different features that we describe in the forthcoming sections.   

\bigskip

To organize the discussion, we start at present time $t=t_0$ and follow the cosmological evolution back in time. The Universe is now nearly flat $H_0^2a_0^2\gg 1$ and it is entering into an accelerating phase. As time recedes, it first goes through a matter dominated era, and then through a radiation dominated era. We assume that non-linearities start to affect the left hand side of \eqref{eq:Friedmann} when the Universe is traversing one of those two epochs. Thus we can write
\begin{equation}
P\left(H^2\right)=\frac {c}{a^{2q}}\,,
\label{eq:Friedmann.flat}
\end{equation}
where $q=2,3/2$  represents radiation or matter dominated eras, and the proportionality constant has been fixed to be $c=\kappa\rho_q^0/3$. 
\section{Singularity analysis}
\subsection{Cosmology close to a regular point}
\label{sec:regular}
Assuming that, as the non-linearities take over, the Hubble paremeter $x=H^2$ is close to any given point $\bar x$ at which the function $P(x)$ is regular, then we can expand the Friedmann  equation \eqref{eq:Friedmann.flat} to obtain  
\begin{equation}
P(\bar x)+ P'(\bar x)\,(H^2-\bar x) =\frac{c}{a^{2q}}\,.
\label{eq:Friedmannregular}
\end{equation}
This expression   can be rewritten as an effective Friedmann equation
\begin{equation}
H^2=\frac{\tilde c}{a^{2q}}+\bar{\bar x} \,,
\label{eq:Friedmannregular.effective}
\end{equation}
with the new constants being $\tilde c=c/P'(\bar x)$ and $\bar {\bar x} =\bar x-P(\bar x)/P'(\bar x)$.
Here we can identify the following possible behaviours:
\begin{itemize}
    \item If the cosmic evolution reaches the non-linear regime when the scale parameter is small enough, then the radiation/matter term dominates over the constant contribution $\bar{\bar x}$, and we get a power law cosmology $a(t)\sim (t-\tilde t)^{1/q}$, where $\tilde t$ is a constant of integration. As the resulting Hubble parameter is not constant $H^2\sim1/(t-\tilde t)^2$, it moves away from $\bar x$ quickly, spoiling the approximation \eqref{eq:Friedmannregular}. For the new value of $H^2$ the solution takes the same form, but with different values for $\tilde c$ and $\tilde t$. This regime can be smoothly reached by receding in time from the present day Universe.
    \item If instead, the Universe reaches the non-linear region with a large scale parameter, equation \eqref{eq:Friedmannregular.effective} would be dominated by its constant part $\bar{\bar x}$, and we get a de Sitter cosmology $a(t)\sim e^{\sqrt{\bar {\bar x}}\,t}$. In this case the Hubble parameter takes a constant value $H^2= \bar{\bar x}$. Consistency with the initial assumption $H^2\sim \bar x$ implies that this regime is only realized when $P(\bar x)\ll P'(\bar x)\,\bar x$. The Hubble parameter remains constant for the subsequent evolution and close to the value $\bar x$, thus the approximation \eqref{eq:Friedmannregular} remains valid. This implies that this solution cannot be reached going backwards in time from present day cosmology.
\end{itemize}
Notice that if we get close to a critical point of $P(x)$, the term $P'(\bar x)(x-\bar x)$ in equation \eqref{eq:Friedmannregular} is replaced by $P^{(p)}(x-\bar x)^p/p!$. The de Sitter behavior does not change, while the power law is instead given by $a(t)\sim (t-\tilde t)^{p/q}$.

\bigskip 

As a conclusion, going back in time from the present day the Universe matches a non-linear regime in which the scale factor behaves as a power law locally in time.

\bigskip 

Notice that the presence of a spatial curvature contribution would modify \eqref{eq:Friedmannregular} as
\begin{equation}
P(\bar x)+ P'(\bar x)\,\left(H^2+\frac k{a^2}-\bar x\right) =\frac{c}{a^{2q}}\,.
\label{eq:Friedmann.regular.curvature}
\end{equation}
This equation can be solved in the two limits stated above. But now the de Sitter regime would be valid only locally in time, since the curvature term would move the argument of $P(x)$ away from $\bar x$ very quickly. In consequence, if the curvature becomes relevant at some time in the evolution, the locally de Sitter regime can also be matched with the present cosmology.  
%
\subsection{Cosmology close to a pole}
\label{sec:pole}
Now suppose that the cosmological evolution drives the spacetime curvature $x=H^2$ close to a pole $\bar x$ of the function $P(x)$. In such a case equation \eqref{eq:Friedmann.flat} takes the form 
\begin{equation}
\frac{r_{p}}{(H^2-\bar x)^{p}}=\frac{c}{a^{2q}}\,.
\label{eq:Friedmann.pole}
\end{equation}
Here $r_p$ is the residue of $P(x)$ around the order $p$ pole at $\bar x$. This can be rearranged as an effective Friedmann equation 
\begin{equation}
H^2=\tilde c\, a^{2q/p}+\bar x\,,
\label{eq:Friedmann.pole.effective}
\end{equation}
where now $\tilde c=({r_{p}}/c)^{1/p}$. We then get two possible behaviours
\begin{itemize}
    \item If the scale factor is small enough as the Hubble parameter $H^2$ reaches the pole, the evolution is that of a de Sitter regime $a(t)\sim e^{\sqrt{\bar x}\,t}$. Then the  Hubble parameter $H^2\sim \bar x$ remains constant and close to the pole, and so the equation \eqref{eq:Friedmann.pole.effective} remains valid. However, the scale factor $a(t)$ grows, until a point in which dominates the right hand side of \eqref{eq:Friedmann.pole.effective} and the de Sitter phase ends. 
    \item If the Hubble parameter reaches the pole $H^2\sim \bar x$ when the scale factor $a(t)$ is large, or if it has grown away from the previously described de Sitter phase, then we get into a regime in which the evolution follows a power law $a(t)\sim (t-\tilde t)^{-p/q}$. Interestingly, the exponent is negative and drives the scale factor down to small values. Whether the cosmology gets then back into the de Sitter phase depends on how much has the Hubble parameter $H^2$ drifted away from $\bar x$ during the power law evolution $H^2\sim 1/(t-\tilde t)^2$. If it is far enough from $\bar x$ and at a regular point, then the approximation in \eqref{eq:Friedmann.pole.effective} is not valid anymore.
\end{itemize}
Again, the presence of curvature would turn the above described de Sitter regime into an approximation which is valid only locally in time, as it happens for the power law regime. 

\bigskip

Remarkably, the above behaviour persists if instead of a pole we have a sharp peak. Indeed, if we write the Friedmann-like equation close to the peak using the regularized expression
\begin{equation}
\frac{r_{p}}{\sqrt{(H^2-\bar x)^{2p}+\epsilon^2}}=\frac{c}{a^{2q}}\,,
\label{eq:Friedmann.peak}
\end{equation}
%
%
%
we can solve for $H^2$ to get en effective Friedmann equation as in \eqref{eq:Friedmann.pole.effective}, containing an additional term in the right hand side of order $\epsilon^2$ proportional to $a^{2q(1-p)/p}$. For a pole of order $p>1$ this contribution will become relevant at early times, when the de Sitter evolution drives the system into small values of $a(t)$, giving rise to a power law evolution.
\bigskip

In conclusion,  we obtained a de Sitter phase which shows up in an evolving cosmology, under the very general requirement of approaching a pole with a small enough scale factor and a small curvature contribution.  It has a natural exit mechanism into a power law cosmology, which can then re-enter the de Sitter phase a few times before leaving it definitively. Alternatively, if we approach a peak, the same behaviour appears, but it does not extends indefinitely to the past. Notice that this behaviour can be smoothly reached by going back in time from the present day Universe.




\section{Discussion}
\begin{figure}
    \centering
    \includegraphics[width=0.99\textwidth]{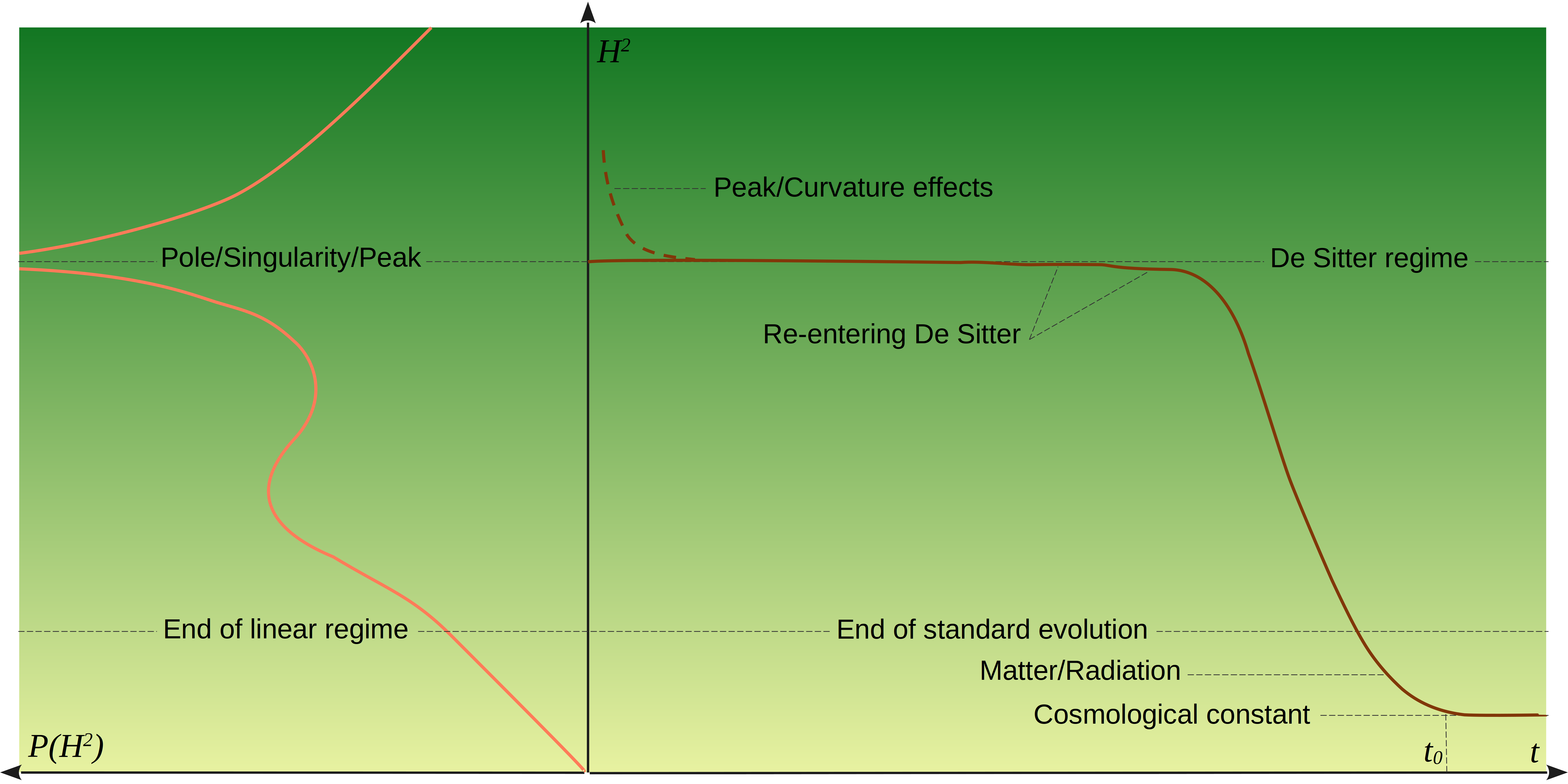}
    \caption{An ilustration of the relation between $P(x)$ and the cosmological evolution. \underline{Left:} a generic form for $P(x)$ as a function of $H^2$ (sideways). \underline{Right:} the corresponding cosmological evolution.}
    \label{fig:evolution}
\end{figure}

To understand the implications of the above results, let us analyze the evolution of the Universe starting from the present state and going backwards in time:
\begin{enumerate}
    \item The function $P(x)$ being dominated by its linear term, the near future, present, and recent past evolution of the Universe follow the standard Friedmann equation. This results in a future de Sitter phase, transitioning as time recedes first into a matter dominated epoch and then into radiation dominated epoch. These are characterized by a power law behavior of the scale factor, which results in a running Hubble parameter $H\sim 1/(t-\tilde t)$.
    \label{step:standard}
    \item At a given point in the past, the Hubble parameter growth turn the non-linear terms of $P(x)$ relevant, getting into a region where we can use the analysis of section \ref{sec:regular}. This cannot happen when the Universe is still diluted enough so as the $\bar x$ term would dominate the right hand side of \eqref{eq:Friedmannregular.effective}, otherwise the resulting de Sitter phase would still running nowadays. The conclusion is that the power law behaviour must persist locally in time, but with a power which is slowly changing as we go to the past. \label{step:runningpowerlaw}
    \item As the Hubble parameter $H\sim 1/(t-\tilde t)$ continues to grow and the scale factor shrinks, we may approach a pole or a sufficiently marked peak of the function $P(x)$. This would result in a de Sitter Universe that would extend into the past. In the case of a pole, if the curvature contribution is still mild this phase would extend indefinitely to the past. If instead the curvature is relevant, or if we are close to a peak, at early times the Universe goes back to a power law evolution. 
    \label{step:deSitter}
    \item Depending on the characteristics of the function $P(x)$, steps \ref{step:runningpowerlaw} and \ref{step:deSitter} can be repeated a number of times as time recedes. 
\end{enumerate}
These results are summarized in Fig.~\ref{fig:evolution}. The general conclusion is that, for very generic forms of the function $P(x)$, the Universe evolves according to a power law with a slowly running power whenever the value of the Hubble parameter is around a regular point of $P(x)$. Such epochs interpolate between succesive de Sitter phases, which appear at the values of the Hubble parameter where the function $P(x)$ has poles or peaks (the de Sitter behavior also shows up at an isolated essential singularity, see Appendix). 

\acknowledgments
We thank Jose Edelstein and Javier Moreno for enlightening comments. This work was partially funded by FONDECYT Regular Grants 1221504, 1200022, 1200293, 1210500, 1210635, and by UNLP  grant X931. A.C. work is partially supported by PRIMUS/23/SCI/005 and GA\v{C}R 22-14791S grants from Charles University.

\appendix

\section{Cosmology close to an isolated essential singularity}

An isolated essential singularity satifies 
\begin{equation}
    \forall p\in \mathbb{Z}_+:P(x)(x-\bar x)^p\to \infty\,.
\end{equation}
To discuss the consequences on the cosmology of such a singularity in the function $P(x)$, we restrict our analysis to the special kind of essential singularities for which the function $P(x)$ close to the singular point $\bar x$ can be written as
\begin{equation}
    P(x)\approx e^{\frac {w_p}{(x-\bar x)^p}}P_p(x)\,,
\end{equation}
where $w_p$ is the weight of the singularity and $P_p(x)$ is a function regular at $\bar x$. The effective Friedmann equation close to the singularity can be then  written in the form
\begin{equation}
H^2=\left({\frac {w_p}{q\log (\tilde c/{a})}}\right)^{1/p}+\bar x\,,
\end{equation}
with the constant $\tilde c=(c/P_p(\bar x))^{1/q}$. 
In this case the de Sitter regime $a(t)\sim e^{\sqrt{\bar 
 x}\,t}$ appears for $a(t)\ll\tilde c$, which in an expanding Universe corresponds to early times. Such phase lasts until the time gets close to $\log\tilde c /\sqrt {\bar x}$ and the first term starts to dominate the equation, moving $H^2$ away from $\bar x$ and ruining the validity of the approximation. This would move the Universe into a regular point, at which the power law regime analyzed in \ref{sec:regular} gets realized.

\bigskip 

As it happened in the other cases, the presence of a spatial curvature term would modify these  considerations, resulting in a power law exit from the de Sitter phase at early times.

 



\end{document}